# Dynamical quantization of the non-Markovian Smoluchowski equation and the quantum tunneling phenomenon

A. O. Bolivar

*Departamento de Física, Universidade Federal de Minas Gerais, Caixa Postal 702, 30123-970 Belo Horizonte, Minas Gerais, Brazil*

Based on the dynamical quantization method we derive a quantum phase-space non-Markovian Smoluchowski equation describing the non-inertial Brownian motion of a harmonic oscillator immersed in a generic environment. In the long-time regime we investigate the tunneling phenomenon by evaluating the quantum Kramers escape rate of a Brownian particle over a potential barrier. As far as a quantum thermal reservoir is concerned, it is found that our steady quantum Kramers rate may depend upon no friction constant in the strong friction domain. This preposterous feature may suggest the existence of non-dissipative quantum tunneling at the low-temperature range, including the zero temperature case. Lastly, we predict that our quantum escape rate non-exponentially decays on the edge of the breakdown of the dissipation-fluctuation relation.

Permanent Address: Instituto Mário Schönberg de Física-Matemática-Filosofia, Ceilândia, D.F., Brazil.

Electronic mail: bolivar@cbpf.br

# I. INTRODUCTION

A material point immersed in an environment undergoes a jittering movement dubbed Brownian motion. For thermal environments the Brownian motion of the tagged particle having position-dependent potential energy $V(x)$ may be described by the Fokker—Planck equation in the absence of inertial force (the so-called Smoluchowski equation) [1]

$$\frac{\partial f(x,t)}{\partial t}=\frac{1}{\gamma}\frac{\partial}{\partial x}\left[\frac{dV(x)}{dx}f(x,t)\right]+\frac{k_BT}{\gamma}\frac{\partial^2 f(x,t)}{\partial x^2} \tag{1}$$

for the time-dependent probability distribution function $f(x,t)$. The diffusion constant, given by the Einstein's fluctuation-dissipation relation [2]

$$D=\frac{k_BT}{\gamma}, \tag{2}$$

is expressed in terms of the friction constant $\gamma$, with dimensions of mass per time, and the thermal energy $k_BT$, where $k_B$ is dubbed the Boltzmann constant and $T$ denotes the heat bath's temperature.

On account of the environmental fluctuations, the escape rate of a Brownian particle over a barrier separating two metastable states located at $x_a$ and $x_b$— in a double-well potential, for instance— is said to be a thermal activation phenomenon [3—9]. On the basis of the Smoluchowski equation (1), Kramers [9] found the following steady-state escape rate

$$\Gamma(\infty)=\frac{m\omega_a\omega_b}{2\pi\gamma}e^{-\frac{V(x_b)}{k_BT}}, \tag{3}$$

where $m$ is the mass of the Brownian particle, $\omega_a$ the oscillation frequency within the bottom well at $x_a$ and $\omega_b$ the oscillation frequency over the barrier located at $x_b$.

It is worth noticing that no diffusive motion, $D=0$, and no escape rate, $\Gamma(\infty)=0$, are predicted at zero temperature. Yet, as far as quantum effects are concerned, metastability phenomenon must be characterized as a quantum tunneling process [8,10—11]. Accordingly, the diffusion coefficient (2) as well as the escape rate (3) cannot vanish at zero-temperature realm.

Because of the wide practical and theoretical interest in chemical physics, for instance, recent studies have been come out showing how quantum effects upon the Smoluchowski equation (1) and the Kramers rate (3) can arise. For example, in the wake of the Caldeira–Leggett approach [12], Ankerhold *et al.* [13—16] have derived a quantum Smoluchowski equation and explored its physical meaning applying it to chemical reactions, mesoscopic physics, and charge transfer in molecules. Meanwhile, further works [11,17,18] have revealed that such a quantum Smoluchowski equation is plagued by some drawbacks, for it may violate the Second Law of Thermodynamics, for instance. Another Hamiltonian account aiming at to quantize the Smoluchowski equation (1) has been developed elsewhere [19,20], whereby a quantum Smoluchowski equation has been derived and the problem of quantum tunneling has been studied at zero temperature, in contrast to Ankerhold *et al.*'s survey. This approach does not depend on the path integral formalism, but it is based on the canonical quantization. Other quantum Smoluchowski equations have been also derived in Refs. [21—25].

Alternatively, we have put forward a quantization method (the so-called dynamical quantization [26—32]) without alluding to the existence of an underlying Hamiltonian model describing the interaction between the tagged particle and the environment. So the main purpose of the present paper is to explore the physical significance of our method of dynamically quantizing the Smoluchowski equation and cope with the problem of tunneling of a Brownian particle immersed in a quantum environment without reckoning with certain assumptions made in our previous work [32] and criticized by Ankerhold *et al.* [37].

Our article is then laid out as follows. In Sect. II we achieve the non-Markovian generalization of the classical Smoluchowski equation in terms of the concept of diffusion energy, whereas in Sec. III we derive a quantum non-Markovian Smoluchowski equation for a harmonic oscillator and find out a steady solution on the basis of which we investigate quantum tunneling in the long-time regime in Sec. IV. At last, summary and perspectives are presented in Sec. V. In addition, some technical details are presented in three appendices.

## II. DERIVING THE NON-MARKOVIAN SMOLUCHOWSKI EQUATION

### A. The Einstein—Langevin stochastic approach

As is well-known, in the classical domain a given isolated system with mass $m$ is described by the position $q = q(t)$ and the linear momentum $\overline{p} = \overline{p}(t) = mdq/dt$, where the parameter $t$ denotes the evolution time and the symbol $d/dt$ a differential operator actuating upon $q$. Its deterministic dynamics fulfill conservative Newton's equations in phase space [33]

$$\frac{d\overline{p}}{dt} = -\frac{dV}{dq}, \tag{4}$$

$$\frac{dq}{dt} = \frac{\overline{p}}{m}, \tag{5}$$

where the inertial force (4) is the position derivative of the potential energy $V = V(q)$ inherent in the particle's motion.

In order to investigate the erratic motion of a tagged material point immersed in an arbitrary environment (a paradigmatic example of open system), Langevin [34] extended the Newton's deterministic approach by putting forward a random generalization of the Newtonian dynamics (4) and (5) through a set of stochastic differential equations (the so-called Langevin equations )

$$\frac{dP}{dt} = -\frac{dV(X)}{dX} - \gamma\frac{P}{m} + \gamma b\Psi(t), \tag{6}$$

$$\frac{dX}{dt} = \frac{P}{m}, \tag{7}$$

where the environmental force,

$$F_{\text{env}}(P,\Psi) = -\gamma\frac{P}{m} + \gamma b\Psi(t), \tag{8}$$

is made up by a *dissipative force* $-\gamma P/m$ — exhibiting no memory effect and accounting for stopping the particle's motion— and by a *fluctuating force*, $\gamma b\Psi(t)$, which is responsible for activating the particle's movement. The parameter $b$ controls such an environmental influence (fluctuations), while the phenomenological parameter $\gamma$, which is expressed in dimensions of mass per time, is termed the damping constant. The term $b\Psi(t)$ in (6) is in dimensions of velocity, i.e., $[length \times time^{-1}]$. Thus, we can readily check that the parameter $b$ may be expressed in dimensions of $[length \times time^{-1/2}]$ and the function $\Psi(t)$ in dimensions of $[time^{-1/2}]$. The existence of the environmental force (8) suggests a relationship between the dissipating and fluctuating processes, i.e., between $b$ and $\gamma$, since $0 < \gamma < \infty$ and $0 < b < \infty$.

From the mathematical viewpoint the quantities $X = X(t)$, $P = P(t)$, and $\Psi = \Psi(t)$ in the Langevin equations (6) and (7) are viewed as random variables in the sense that there exists a probability distribution function, $\mathcal{F}_{XP\Psi}(x,p,\psi,t)$, associated with the stochastic system $\{X,P,\Psi\}$, expressed in terms of the possible values $x = \{x_i\}$, $p = \{p_i\}$, and $\psi = \{\psi_i\}$, with $i \geq 1$, distributed about the sharp values $q$, $\overline{p}$ and $\varphi$ of $X$, $P$, and $\Psi$, respectively.

In the Einstein—Langevin randomization process the average values of $X$, $P$ and $\Psi$ are expressed as

$$\langle X \rangle = \iiint\limits_{-\infty}^{+\infty} x\mathcal{F}_{XP\Psi}(x,p,\psi,t)dxdpd\psi, \tag{9}$$

$$\langle P \rangle = \iiint\limits_{-\infty}^{+\infty} p\mathcal{F}_{XP\Psi}(x,p,\psi,t)dxdpd\psi, \tag{10}$$

$$\langle \Psi \rangle = \iiint\limits_{-\infty}^{+\infty} \psi\mathcal{F}_{XP\Psi}(x,p,\psi,t)dxdpd\psi. \tag{11}$$

In the deterministic case, i.e., as $b = 0$ and $\mathcal{F}_{XP\Psi}(x,p,\psi,t) = \delta(x-q)\delta(p-\overline{p})\delta(\psi-\varphi)$, the deterministic variables $x = q$ and $\overline{p} = p = mdx/dt$ are recovered as the respective average values $\langle X \rangle = q$ and $\langle P \rangle = \overline{p}$.

## 1. The generalized Smoluchowski equation

We start from the Langevin equations (6) and (7) in the Smoluchowski limit (also known as the large friction case)

$$|F_{\text{env}}(P,\Psi)| \gg \left|\frac{dP}{dt}\right|, \tag{12}$$

such that inertial effects in (6) can be negligible, i.e., $dP/dt = 0$. So the focus on the motion of the Brownian particle is gained through a single stochastic differential equation given by

$$\gamma\frac{dX}{dt} = -\frac{dV(X)}{dX} + \gamma b\Psi(t), \tag{13}$$

in which the dissipative force $\gamma dX/dt$ is offset by the conservative force, $-dV(X)/dX$, and the stochastic force, $\gamma b\Psi(t)$.

Noticing that

$$\lim_{\varepsilon \to 0} \int_{t}^{t+\varepsilon} \langle \Psi(t') \rangle dt' = 0, \tag{14}$$

the Langevin equation (13), divided by $\gamma$, gives rise to the following Fokker—Planck equation on configuration space in the Gaussian approximation (see Appendix A)

$$\frac{\partial f(x,t)}{\partial t} = \frac{\partial}{\partial x}\left[\frac{1}{\gamma}\frac{dV(x)}{dx} - b\langle \Psi(t) \rangle\right] f(x,t) + \mathcal{D}(t)\frac{\partial^2 f(x,t)}{\partial x^2}, \tag{15}$$

where the mean value of $\Psi(t)$ is given by

$$\langle \Psi(t) \rangle = \lim_{\varepsilon \to 0} \frac{1}{\varepsilon} \int_{t}^{t+\varepsilon} \langle \Psi(t') \rangle dt', \tag{16}$$

and the time-dependent diffusion coefficient

$$\mathcal{D}(t) = \frac{\mathcal{E}(t)}{\gamma} \tag{17}$$

is expressed in terms of the following function

$$\mathcal{E}(t) = \frac{\gamma b^2}{2} I(t) = \frac{\gamma b^2}{2} \lim_{\varepsilon \to 0} \frac{1}{\varepsilon} \iint_{t}^{t+\varepsilon} \langle \Psi(t^{'}) \Psi(t^{''}) \rangle dt^{'} dt^{''}, \tag{18}$$

which has dimensions of energy, i.e., $[mass \times length^2 \times time^{-2}]$. Hence we call $\mathcal{E}(t)$ the diffusion energy responsible for the Brownian motion of the particle dipped in a generic environment.

The time-dependent function $I(t)$ in (18), which encapsulates all the fluctuation effects on the Brownian particle through the auto-correlation function $\langle b\Psi(t') b\Psi(t'') \rangle$, is a dimensionless term because $b$ and $\Psi(t)$ in (6) has been assumed to have dimensions of $[length \times time^{-1/2}]$ and $[time^{-1/2}]$, respectively.

An outstanding feature underlying our diffusion energy concept in (17) is that $\mathcal{E}(t) = \gamma \mathcal{D}(t)$ conveys in tandem the fluctuation phenomenon through the diffusion coefficient $\mathcal{D}(t)$ as well as the dissipation process encapsulated in the friction constant $\gamma$. We can then state that our definition of diffusion energy (17), which fulfils the validity condition

$$0 < \mathcal{E}(t) < \infty, \tag{19}$$

stands out a general fluctuation—dissipation relation underlying all open systems described by the Langevin equation (13) and its corresponding Fokker—Planck equation (15). Both cases $\mathcal{E}(t) = 0$ and $\mathcal{E}(t) = \infty$ are not concerned, for they may violate the fluctuation—dissipation relation. The former case may lead to dissipation without fluctuation, while the latter one may give rise to fluctuation without dissipation.

In the long-time regime we assume that the diffusion energy (18) renders steady, i.e.,

$$\lim_{t\to\infty} \mathcal{E}(t) = \mathcal{E}(\infty) \tag{20}$$

with

$$\lim_{t\to\infty} I(t) = I(\infty) = 1. \tag{21}$$

The physical significance of condition (21) has to do with the fact that the environmental fluctuations do possess Markovian correlations, i.e., $I(t)$ displays a local (short) range behavior decaying to one in the steady regime. In generic open systems non-Markovian effects show up at the range $0 < t < \infty$. The function $I(t)$ exhibiting the asymptotic behavior (21) can be built up as (see Appendix B)

$$I(t) = 1 - e^{-\frac{t}{t_c}}, \tag{22}$$

such that at $t = 0$ the diffusion energy associated with the initial probability distribution function $f(x, t = 0) = \delta(x)$ vanishes: $\mathcal{E}(0) = 0$. In (22), $t_c$ denotes the correlation time making the fluctuations non-Markovian.

Accordingly, from (18) we obtain the fluctuation-dissipation relation in the form

$$b = \sqrt{\frac{2\mathcal{E}(\infty)}{\gamma}} \tag{23}$$

and the Fokker—Planck equation (15) reads

$$\frac{\partial f(x,t)}{\partial t} = \frac{1}{\gamma}\frac{\partial}{\partial x}\left[\frac{\partial \mathcal{V}(x,t)}{\partial x} f(x,t)\right] + \frac{\mathcal{E}(\infty)}{\gamma} I(t) \frac{\partial^2 f(x,t)}{\partial x^2}, \tag{24}$$

with the effective potential

$$\mathcal{V}(x,t) = V(x) - x\sqrt{2\gamma\mathcal{E}(\infty)}\langle\Psi(t)\rangle. \tag{25}$$

For the case in which $\langle\Psi(t)\rangle = 0$ and $I(t) = 1$, the Fokker—Planck equation (24) reduces to the Markovian Smoluchowski equation [1]

$$\frac{\partial f(x,t)}{\partial t} = \frac{1}{\gamma}\frac{\partial}{\partial x}\left[\frac{dV(x)}{dx} f(x,t)\right] + \frac{\mathcal{E}(\infty)}{\gamma}\frac{\partial^2 f(x,t)}{\partial x^2}. \tag{26}$$

Thereby, we dub Eq. (24) the generalized Smoluchowski equation because it reckons with both non-Markovian and averaging effects.

# III. DYNAMICAL QUANTIZATION OF OUR NON-MARKOVIAN SMOLUCHOWSKI EQUATION

## A. Quantum non-thermal open systems

Our classical Smoluchowski equation (24) with (22) at point $x^{'}$ for the case of a harmonic oscillator, $V(x') = kx'^2/2$, where the constant $k$ has dimensions of $[mass \times time^{-2}]$, reads

$$\frac{\partial f(x,t)}{\partial t} = \frac{k}{\gamma}\frac{\partial}{\partial x}[xf(x,t)] + \frac{\mathcal{E}(\infty)}{\gamma}\left(1 - e^{-\frac{t}{t_c}}\right)\frac{\partial^2 f(x,t)}{\partial x^2}, \tag{27}$$

in terms of the position

$$x = x^{'} - \frac{1}{k}\sqrt{2\gamma\mathcal{E}(\infty)}\langle\Psi(t)\rangle. \tag{28}$$

Our purpose is to quantize the non-Markovian Smoluchowski equation (27). To this end, we start with it at points $x_1$ and $x_2$, i.e.,

$$\frac{\partial\chi(x_1,t)}{\partial t} = \frac{k}{\gamma}\chi(x_1,t) + \frac{kx_1}{\gamma}\frac{\partial\chi(x_1,t)}{\partial x_1} + \frac{\overline{\mathcal{E}}(t)}{\gamma}\frac{\partial^2\chi(x_1,t)}{\partial x_1^2} \tag{29}$$

and

$$\frac{\partial\chi(x_2,t)}{\partial t} = \frac{k}{\gamma}\chi(x_2,t) + \frac{kx_2}{\gamma}\frac{\partial\chi(x_2,t)}{\partial x_2} + \frac{\overline{\mathcal{E}}(t)}{\gamma}\frac{\partial^2\chi(x_2,t)}{\partial x_2^2}\,. \tag{30}$$

Solutions $\chi(x_1,t)$ and $\chi(x_2,t)$ are deemed to be associated with the diffusion energy $\overline{\mathcal{E}}(t) = 2\mathcal{E}(\infty)(1 - e^{-t/t_c})$. By multiplying (29) by $\chi(x_2,t)$ and (30) by $\chi(x_1,t)$ and adding the resulting equations, we obtain

$$\frac{\partial\xi}{\partial t} = \frac{2k}{\gamma}\xi + \frac{k}{\gamma}\left(x_1\frac{\partial\xi}{\partial x_1} + x_2\frac{\partial\xi}{\partial x_2}\right) + \frac{2}{\gamma}\mathcal{E}(\infty)\left(1 - e^{-\frac{t}{t_c}}\right)\left(\frac{\partial^2\xi}{\partial x_1^2} + \frac{\partial^2\xi}{\partial x_2^2}\right), \tag{31}$$

where $\xi \equiv \xi(x_1,x_2,t) = \chi(x_1,t)\chi(x_2,t) = \sqrt{f(x_1,t)f(x_2,t)}$.

By performing the following change of variables into configuration space, $(x_1,x_2) \longmapsto (x,\eta\hbar)$,

$$x_1 = x - \frac{\eta\hbar}{2}, \tag{32}$$

$$x_2 = x + \frac{\eta\hbar}{2}, \tag{33}$$

and making use of the relations

$$\frac{\partial}{\partial x_1} = \frac{1}{2}\frac{\partial}{\partial x} - \frac{1}{\hbar}\frac{\partial}{\partial \eta} \qquad ; \qquad \frac{\partial}{\partial x_2} = \frac{1}{2}\frac{\partial}{\partial x} + \frac{1}{\hbar}\frac{\partial}{\partial \eta},$$

the classical equation (31) changes into the quantum equation of motion

$$\frac{\partial \rho}{\partial t} = \frac{2k}{\gamma}\rho + \frac{k}{\gamma}x\frac{\partial \rho}{\partial x} + \frac{k}{\gamma}\eta\frac{\partial \rho}{\partial \eta} + \frac{\mathcal{E}_\hbar(t)}{\gamma}\left(\frac{\partial^2 \rho}{\partial x^2} + \frac{4}{\hbar^2}\frac{\partial^2 \rho}{\partial \eta^2}\right), \tag{34}$$

where we have replaced the solution $\xi = \xi(x_1, x_2, t)$ with $\rho = \rho(x, \eta, t)$ and employed the notation

$$\mathcal{E}_\hbar(t) = \gamma \mathcal{D}_\hbar(t) \; = \mathcal{E}_\hbar(\infty)\left(1 - e^{-\frac{t}{t_c}}\right) \tag{35}$$

to display the quantum nature of the diffusion energy which turns out to be now expressed in terms of the Planck constant ħ. Notice that the constants $k, \gamma$, and the dimensionless function $I(t)$ hold classical during the classical-quantum transition (32) and (33).

The geometric meaning of our quantization conditions (32) and (33) is related to the existence of a minimal distance between $x_1$ and $x_2$ due to the quantum nature of space, $x_2 - x_1 = \eta\hbar$, such that in the classical limit $\hbar \to 0$, physically interpreted as

$$|\eta\hbar| \ll |x_2 - x_1|, \tag{36}$$

the result $x_2 = x_1 = x$ can be recovered. Accordingly, the classical limit of the quantum equation of motion (34) leads rightly to our Smoluchowski equation (27) because

$$\lim_{\hbar\to 0} \rho(x, \eta, t) = \lim_{\hbar\to 0} \xi(x_1, x_2, t) = \sqrt{f(x,t)f(x,t)} = f(x,t), \tag{37}$$

provided that

$$\lim_{\hbar\to 0} \mathcal{E}_\hbar(t) = \mathcal{E}(t). \tag{38}$$

By making use of the Fourier transform

$$W(x, p, t) = \frac{1}{2\pi}\int_{-\infty}^{\infty} \rho(x, \eta, t) e^{ip\eta}\, d\eta, \tag{39}$$

which changes the variables from quantum configuration space $(x, \eta ħ)$ onto quantum phase space $(x, p; ħ)$, equation of motion (33) turns out to be written down as

$$\frac{\partial W}{\partial t} = \frac{k}{\gamma} x \frac{\partial W}{\partial x} - \frac{k}{\gamma} p \frac{\partial W}{\partial p} + \frac{\mathcal{E}_ħ(t)}{\gamma} \frac{\partial^2 W}{\partial x^2} + \left[ \frac{k}{\gamma} - \frac{4\mathcal{E}_ħ(t)}{ħ^2 \gamma} p^2 \right] W \tag{40}$$

in terms of the function $W \equiv W(x, p, t)$. Because the exponential factor $e^{ip\eta}$ in (39) is to be a dimensionless term, it follows that the variable $p$ is to have dimensions of linear momentum.

On account of the following property (see Appendix C)

$$\lim_{ħ \to 0} W(x, p, t) = \mathcal{F}(x, p, t) = f(x, t)\delta(p), \tag{41}$$

the classical limit of (40) leads to the Smoluchowski equation (27)as well.

### *1. The initial condition*

Our quantum phase-space Smoluchowski equation (40) may be solved starting from the non-thermal initial condition

$$W(x, p, t = 0) = \frac{1}{\pi ħ} e^{-\left(\frac{ap^2}{ħ} + \frac{x^2}{ħ a}\right)} \tag{42}$$

leading to the distribution $F(x, p) = \delta(x)\delta(p)$ in the classical limit, $ħ \to 0$. The constant $a$ has dimensions of time per mass, that is, we may take $a = 1/m\omega$, where $\omega = \sqrt{m/k}$ is the oscillation frequency of the harmonic oscillator. The probability distribution function (42) generates the fluctuations

$$\langle P^2 \rangle = \frac{ħ m\omega}{2}, \tag{43}$$

$$\langle X^2 \rangle = \frac{ħ}{2m\omega}, \tag{44}$$

which are compatible with the Heisenberg indeterminacy principle

$$\sqrt{\langle P^2 \rangle \langle X^2 \rangle} = \frac{ħ}{2}. \tag{45}$$

The initial mean quantum-mechanical energy of a Brownian harmonic oscillator is given by

$$\langle \mathrm{E}(0) \rangle = \frac{\langle P^2 \rangle}{2m} + \frac{m\omega^2}{2} \langle X^2 \rangle = \frac{ħ\omega}{2}, \tag{46}$$

albeit its initial diffusion energy (35) is null, i.e., $\mathcal{E}_ħ(0) = 0$.

*2. The time-dependent solution*

A time-dependent solution to (40) reads

$$W(x,p,t) = \frac{1}{\pi\hbar} e^{-\left[\frac{x^2}{4B_\hbar(t)} + \frac{4B_\hbar(t)p^2}{\hbar^2}\right]}, \tag{47}$$

where the function

$$B_\hbar(t) = e^{-\frac{2kt}{\gamma}}\left[\frac{\hbar}{4m\omega} - \mathcal{E}_\hbar(\infty)\left(\frac{1}{2k} + \frac{t_c}{(\gamma - 2kt_c)}\right)\right] + \mathcal{E}_\hbar(\infty)\left(\frac{1}{2k} + \frac{t_c e^{-\frac{t}{t_c}}}{(\gamma - 2kt_c)}\right) \tag{48}$$

is expressed in terms of the evolution time $t$, the relaxation time $t_r = \gamma/k$, and the correlation time $t_c$ responsible for non-Markovian effects.

*3. The long-time regime*

In the long-time regime, $t \to \infty$, physically interpreted as $t \gg t_c, t_r$, the probability distribution function (47) renders steady, that is,

$$W_{\rm st}(x,p) = \frac{1}{\pi\hbar} e^{-\left[\frac{kx^2}{2\mathcal{E}_\hbar(\infty)} + \frac{2\mathcal{E}_\hbar(\infty)p^2}{k\hbar^2}\right]} \tag{49}$$

leading to the fluctuations

$$\langle P^2\rangle = \frac{k\hbar^2}{4\mathcal{E}_\hbar(\infty)}, \tag{50}$$

$$\langle X^2\rangle = \frac{\mathcal{E}_\hbar(\infty)}{k}, \tag{51}$$

compatible with the Heisenberg relation (45), and the steady mean quantum-mechanical energy

$$\langle \mathrm{E}(\infty)\rangle = \frac{k\hbar^2}{8m\mathcal{E}_\hbar(\infty)} + \frac{\mathcal{E}_\hbar(\infty)}{2}, \tag{52}$$

which is to be defined for the quantum diffusion energy satisfying the validity range $0 < \mathcal{E}_\hbar(\infty) < 0$. In the especial cases $\mathcal{E}_\hbar(\infty) = 0$ and $\mathcal{E}_\hbar(\infty) = \infty$ expression (52) blows up, thereby violating the quantum dissipation-fluctuation relation.

**B. Quantum thermal open systems**

Imagine that the environment is made up of a set of harmonic oscillators having the same oscillation frequency $\omega$, so that the mean energy of this quantum heat bath after attaining the thermodynamic equilibrium at temperature $T$ is given by [35]

$$\overline{E} = \frac{\omega\hbar}{2}\left(\frac{e^{\frac{\omega\hbar}{k_B T}}+1}{e^{\frac{\omega\hbar}{k_B T}}-1}\right) = \frac{\omega\hbar}{2}\coth\left(\frac{\omega\hbar}{2k_B T}\right), \tag{53}$$

where the ħ-dependent energy, $\omega\hbar/2$, corresponds to the mean quantum-mechanical energy (46) of the heat bath prepared at the initial state (42) and $k_B T$ the classical thermal energy of the heat bath, where $k_B$ is the Boltzmann constant.

We now assume that the Brownian particle's quantum diffusion energy $\mathcal{E}_\hbar(\infty)$ can be identified with the heat bath's quantum thermal energy $\overline{E}$ , i.e.,

$$\mathcal{E}_\hbar(\infty) = \frac{\omega\hbar}{2}\coth\left(\frac{\omega\hbar}{2k_B T}\right). \tag{54}$$

Using (35) in the steady regime the quantum diffusion coefficient can be derived as

$$\mathcal{D}_\hbar(\infty) = \frac{\omega\hbar}{2\gamma}\coth\left(\frac{\omega\hbar}{2k_B T}\right). \tag{55}$$

The steady probability distribution function (49) becomes the steady quantum-mechanical equilibrium function

$$W_{\mathrm{st}}(x,p) = \frac{1}{\pi\hbar} e^{\frac{-m\omega x^2}{\hbar\coth\left(\frac{\omega\hbar}{2k_B T}\right)} - \frac{1}{m\omega\hbar}\coth\left(\frac{\omega\hbar}{2k_B T}\right)p^2} \tag{56}$$

leading to the fluctuations (50) and (51) in the form

$$\langle P^2\rangle = \frac{m\omega\hbar}{2\coth\left(\frac{\omega\hbar}{2k_B T}\right)}, \tag{57}$$

$$\langle X^2\rangle = \frac{\hbar}{2m\omega}\coth\left(\frac{\omega\hbar}{2k_B T}\right). \tag{58}$$

The mean quantum-mechanical energy (52) is then given by

$$\langle \mathrm{E}(\infty)\rangle = \frac{\omega\hbar}{4\coth\left(\frac{\omega\hbar}{2k_B T}\right)} + \frac{\omega\hbar}{4}\coth\left(\frac{\omega\hbar}{2k_B T}\right). \tag{59}$$

*1. Zero temperature*

The steady probability distribution function (56) at zero temperature reads

$$W_{\mathrm{st}}^{(T=0)}(x,p) = \frac{1}{\pi\hbar} e^{-\left[\frac{m\omega\, x^2}{\hbar}+\frac{p^2}{m\omega\,\hbar}\right]} \tag{60}$$

and the fluctuations (57) and (58) become

$$\langle P^2\rangle^{(T=0)} = \frac{m\omega\hbar}{2}, \tag{61}$$

$$\langle X^2\rangle^{(T=0)} = \frac{\hbar}{2m\omega}. \tag{62}$$

The mean quantum-mechanical energy (59) is then given by

$$\langle \mathrm{E}(\infty)\rangle^{(T=0)} = \frac{\omega\hbar}{2} \tag{63}$$

which equals the initial energy (46) and the diffusion energy (54) in the steady regime

$$\mathcal{E}_{\hbar}^{(T=0)}(\infty) = \frac{\omega\hbar}{2}, \tag{64}$$

whereas the quantum diffusion coefficient (55) turns out to be

$$\mathcal{D}_{\hbar}^{(T=0)}(\infty) = \frac{\omega\hbar}{2\gamma}. \tag{65}$$

On account of the energy conservation (54), results (60—65) provide physical significance to our quantum phase-space Smoluchowski equation (40) at zero temperature.

*2. The classical limit*

The classical limit $\hbar \to 0$ should be physically interpreted as the classical thermal energy is deemed to be too large in comparison to the quantal energy, i.e.,

$$k_B T \gg \frac{\omega\hbar}{2}. \tag{66}$$

Thus, our quantum Smoluchowski equation (40) reproduces the classical Smoluchowski equation (27) at the high temperature range, whereas the classical limit of (59) leads rightly to the principle of equipartition of energy

$$\lim_{\hbar\to 0}\langle \mathrm{E}(\infty)\rangle = \frac{k_B T}{2} = \langle E(\infty)\rangle, \tag{67}$$

while the quantum diffusion energy (54) turns out to be the thermal energy

$$\mathcal{E}(\infty) = k_B T. \tag{68}$$

We also obtain the Einstein's celebrated relation

$$\mathcal{D}(\infty) = \frac{k_B T}{\gamma} \tag{69}$$

as quantum effects can be negligible in the diffusion constant (55). In other words, the Einstein diffusion constant (69) maintains valid at high temperature $T \gg \omega\hbar/2k_B$.

Furthermore, we can derive the Maxwell—Boltzmann distribution function

$$f(x) = \frac{1}{\sqrt{2\pi k_B T}} e^{-\frac{kx^2}{2k_B T}}. \tag{70}$$

as the classical limit of the (marginal) probability distribution function (56).

In view of the energy conservation (54), results (66—70) yield physical meaning to our quantum phase-space Smoluchowski equation (40) at high temperature by reproducing the well-known tenets of the classical theory of non-inertial Brownian motion at thermodynamic equilibrium.

## IV. AN APPLICATION: THE QUANTUM KRAMERS EQUATION

### A. Quantum non-thermal open systems

In order to calculate the quantum Kramers escape rate of a Brownian particle over a potential barrier in the strong friction regime, we consider as a starting point the steady solution (49), i.e.,

$$W_{\mathrm{st}}(x,p) = \frac{1}{\pi\hbar} e^{-\left[\frac{V(x)}{\mathcal{E}_\hbar(\infty)} + \frac{2\mathcal{E}_\hbar(\infty)p^2}{k\hbar^2}\right]}. \tag{71}$$

We perform an expansion of the potential function $V(x) = kx^2/2$ in a Taylor series around a certain point $x_0$, i.e., $V(x) \sim V(x_0) + (k/2)(x - x_0)^2$, and construct the Markovian non-equilibrium function

$$W_{\mathrm{neq}}(x,p) = \alpha\varphi(x,p) e^{-\frac{\left[V(x_0) + \frac{k}{2}(x-x_0)^2\right]}{\mathcal{E}_\hbar(\infty)} - \frac{2\mathcal{E}_\hbar(\infty)}{\hbar^2 k}p^2}, \tag{72}$$

$\alpha$ being a constant and $\varphi = \varphi(x,p)$ a function to be determined. On inserting solution (72) into our quantum Smoluchowski equation (40), given by

$$\frac{k}{\gamma}(x-x_0)\frac{\partial W_{\text{neq}}}{\partial x}-\frac{k}{\gamma}p\frac{\partial W_{\text{neq}}}{\partial p}+\frac{\mathcal{E}_\hbar(\infty)}{\gamma}\frac{\partial^2 W_{\text{neq}}}{\partial x^2}+\left[\frac{k}{\gamma}-\frac{4\mathcal{E}_\hbar(\infty)}{\gamma\hbar^2}p^2\right]W_{\text{neq}}=0, \quad (73)$$

with $W_{\text{neq}} = W_{\text{neq}}(x,p)$, we obtain the equation of motion

$$\frac{k}{\gamma}(x-x_0)\frac{\partial \varphi}{\partial x}-\frac{k}{\gamma}p\frac{\partial \varphi}{\partial p}+\frac{\mathcal{E}_\hbar(\infty)}{\gamma}\frac{\partial^2 \varphi}{\partial x^2}-\frac{2k}{\gamma}(x-x_0)\frac{\partial \varphi}{\partial x}=0, \quad (74)$$

which changes into

$$\frac{d^2\varphi}{d\xi^2}=\frac{k}{\beta^2\mathcal{E}_\hbar(\infty)}\xi\frac{d\varphi}{d\xi} \quad (75)$$

in terms of the variable

$$\xi=\beta(x-x_0)-p. \quad (76)$$

The constant $\beta$ in (75) and (76) has dimensions of $[mass \times time^{-1}]$ such that the term $\beta(x-x_0)$ in (76) is in dimensions of momentum, like $p$ and $\xi$. A solution to the differential equation (75) reads

$$\varphi(\xi)=\left(\frac{-k}{2\pi\beta^2\mathcal{E}_\hbar(\infty)}\right)^{1/2}\int_{-\infty}^{\xi} e^{\frac{k\xi^2}{2\beta^2\mathcal{E}_\hbar(\infty)}}d\xi \;, \quad (77)$$

fulfilling the condition $\varphi(\xi\to\infty)=1$. This result (77) is only possible if the potential curvature is negative, i.e., $k<0$. Hence, we hereafter use $x_0\equiv x_b$, and $k=-m\omega_b^2$, where the quantity $\omega_b$ denotes the particle's oscillation frequency over the potential barrier at $x_b$. So the barrier top is located at point $x_b$, while the two bottom wells are at $x_a$ and $x_c$, such that $V(x_c)=0=V(x_a)$, with $x_a < x_b$.

Using (77), the stationary function (72) becomes

$$W_{\text{neq}}(x,p)=\alpha\left(\frac{m\omega_b^2}{\pi\beta^2\mathcal{E}_\hbar(\infty)}\right)^{\frac{1}{2}} e^{-\frac{\left[V(x_b)-\frac{m\omega_b^2}{2}(x-x_b)^2\right]}{\mathcal{E}_\hbar(\infty)}+\frac{2\mathcal{E}_\hbar(\infty)}{\hbar^2 m\omega_b^2}p^2}\int_{-\infty}^{\xi} e^{-\frac{m\omega_b^2\xi^2}{2\beta^2\mathcal{E}_\hbar(\infty)}}d\xi \quad (78)$$

from which we can find out the probability current as

$$J_b=\frac{1}{m}\int_{-\infty}^{\infty} pW_{\text{neq}}(x=x_b,p)\,dp=-\frac{\alpha\omega_b^4\hbar^3 m}{4\mathcal{E}_\hbar(\infty)}\frac{e^{-\frac{V(x_b)}{\mathcal{E}_\hbar(\infty)}}}{\sqrt{m^2\hbar^2\omega_b^4-4\beta^2\mathcal{E}_\hbar^2(\infty)}}, \quad (79)$$

where we used the result

$$\int_{-\infty}^{\infty} e^{-up^2}\,pdp\int_{-\infty}^{\xi=p} e^{-v\xi^2}d\xi=\frac{1}{2u}\left(\frac{\pi}{u+v}\right)^{1/2},$$

$u$ and $v$ being non-negative constants.

The number of particles through the well located at $x_a$ is calculated as

$$n_a = \iint_{-\infty}^{\infty} W_{\text{eq}}(x,p)dpdx = \alpha\pi\hbar, \tag{80}$$

whereby we used the equilibrium distribution function

$$W_{\text{eq}}(x,p) = \alpha e^{-\frac{2\mathcal{E}_\hbar(\infty)p^2}{\hbar^2 m\omega_a^2} - \frac{m\omega_a^2}{2\mathcal{E}_\hbar(\infty)}(x-x_a)^2} \tag{81}$$

coming from the non-equilibrum function (72) for $\varphi(x,p) = 1$ calculated at the well $x_a$, with $k_a = m\omega_a^2$.

Making use of (79) and (80) the quantum Kramers escape rate of a Brownian particle immersed in a non-thermal environment is then written down as

$$\Gamma(\infty) = \frac{|J_b|}{n_a} = \frac{\omega_b^4\hbar^2 m}{4\pi\mathcal{E}_\hbar(\infty)} \frac{e^{-\frac{V(x_b)}{\mathcal{E}_\hbar(\infty)}}}{\sqrt{m^2\hbar^2\omega_b^4 - 4\beta^2\mathcal{E}_\hbar^2(\infty)}} \tag{82}$$

in terms of the steady quantum diffusion energy, $\mathcal{E}_\hbar(\infty) = \gamma b_\hbar^2/2 = \gamma\mathcal{D}_\hbar(\infty)$, that fulfills the condition

$$0 < \mathcal{E}_\hbar(\infty) < \frac{m\hbar\omega_b^2}{2\beta}. \tag{83}$$

It is worth stressing that our main upshot (82) does display dimensions of inverse of time, as it should be expected, as well as being compatible with the dissipation-fluctuation relation $\mathcal{E}_\hbar(\infty)$.

Our quantum escape rate (82) decays exponentially with respect to the ratio$-V(x_b)/\mathcal{E}_\hbar(\infty)$. In addition, the so-called Arrhenius factor, given by

$$\sigma = \frac{\omega_b^4\hbar^2 m}{4\pi\mathcal{E}_\hbar(\infty)} \frac{1}{\sqrt{m^2\hbar^2\omega_b^4 - 4\beta^2\mathcal{E}_\hbar^2(\infty)}}, \tag{84}$$

does depend upon the arbitrary parameter $\beta$ introduced by the change of variables (76). Because it is expressed in dimensions of mass per time, we may consider the following cases: $\beta = \gamma$ and $\beta = Nm\omega_b$, where $N$ is a dimensionless number such that $0 \leq N < \infty$.

#### *1. The* $\boldsymbol{\beta = Nm\omega_b}$ *case*

Considering $\beta = Nm\omega_b$ our quantum rate (82) turns out to be given by

$$\Gamma(\infty) = \frac{\omega_b^3 \hbar^2}{4\pi \mathcal{E}_\hbar(\infty)} \frac{e^{-\frac{V(x_b)}{\mathcal{E}_\hbar(\infty)}}}{\sqrt{\hbar^2 \omega_b^2 - 4N^2 \mathcal{E}_\hbar^2(\infty)}} \tag{85}$$

with

$$0 \leq N < \frac{\hbar\omega_b}{2\mathcal{E}_\hbar(\infty)}. \tag{86}$$

Taking $N = 0$ in (85) leads to

$$\Gamma(\infty) = \frac{\omega_b^2 \hbar}{4\pi \mathcal{E}_\hbar(\infty)} e^{-\frac{V(x_b)}{\mathcal{E}_\hbar(\infty)}}. \tag{87}$$

#### 2*. The* $\boldsymbol{\beta = \gamma}$ *case*

For $\beta = \gamma$, our rate (82) reads

$$\Gamma(\infty) = \frac{\omega_b^4 \hbar^2 m}{4\pi \mathcal{E}_\hbar(\infty)} \frac{e^{-\frac{V(x_b)}{\mathcal{E}_\hbar(\infty)}}}{\sqrt{m^2\hbar^2 \omega_b^4 - 4\gamma^2 \mathcal{E}_\hbar^2(\infty)}} \tag{88}$$

provided that

$$0 < \mathcal{E}_\hbar(\infty) < \frac{\omega_b^2 \hbar m}{2\gamma}. \tag{89}$$

### B. Quantum thermal open systems

For thermal systems the quantum diffusion energy $\mathcal{E}_\hbar(\infty)$ given by (54) is evaluated over the potential barrier, i.e.,

$$\mathcal{E}_\hbar(\infty) = \frac{\omega_b \hbar}{2} \coth\left(\frac{\omega_b \hbar}{2k_B T}\right). \tag{90}$$

Accordingly, our quantum Kramers rate (82) reads

$$\Gamma(\infty) = \frac{m\omega_b^2}{2\pi \coth\left(\frac{\omega_b \hbar}{2k_B T}\right)} \frac{e^{-\frac{2V(x_b)}{\omega_b \hbar \coth\left(\frac{\omega_b \hbar}{2k_B T}\right)}}}{\sqrt{m^2\omega_b^2 - \beta^2 \coth^2\left(\frac{\omega_b \hbar}{2k_B T}\right)}}. \tag{91}$$

valid for the low-temperature regime

$$0 < \coth\left(\frac{\omega_b \hbar}{2k_B T}\right) < \frac{m\omega_b}{\beta}.$$

At zero temperature, Equation (90) becomes $\mathcal{E}_{\hbar}(\infty) = \omega\hbar/2$ and (91) reduces to

$$\Gamma^{(T=0)}(\infty) = \frac{m\omega_b^2}{2\pi} \frac{e^{-\frac{2V(x_b)}{\omega_b\hbar}}}{\sqrt{m^2\omega_b^2 - \beta^2}}. \tag{92}$$

### *1. The $\boldsymbol{\beta = Nm\omega_b}$ case*

For $\beta = Nm\omega_b$, we find from (91) the following quantum escape rate

$$\Gamma(\infty) = \frac{\omega_b}{2\pi\coth\left(\frac{\omega_b\hbar}{2k_BT}\right)} \frac{e^{-\frac{2V(x_b)}{\omega_b\hbar\coth\left(\frac{\omega_b\hbar}{2k_BT}\right)}}}{\sqrt{1 - N^2\coth^2\left(\frac{\omega_b\hbar}{2k_BT}\right)}} \tag{93}$$

with

$$0 \le N < \frac{1}{\coth\left(\frac{\omega_b\hbar}{2k_BT}\right)}. \tag{94}$$

The $N = 0$ case provides

$$\Gamma(\infty) = \frac{\omega_b}{2\pi\coth\left(\frac{\omega_b\hbar}{2k_BT}\right)} e^{-\frac{2V(x_b)}{\omega_b\hbar\coth\left(\frac{\omega_b\hbar}{2k_BT}\right)}} \tag{95}$$

At zero temperature, (93) reduces to

$$\Gamma^{(T=0)}(\infty) = \frac{\omega_b}{2\pi} \frac{e^{-\frac{2V(x_b)}{\omega_b\hbar}}}{\sqrt{1 - N^2}} \tag{96}$$

with $0 \le N < 1$. As $N = 0$ we have

$$\Gamma^{(T=0)}(\infty) = \frac{\omega_b}{2\pi} e^{-\frac{2V(x_b)}{\omega_b\hbar}}. \tag{97}$$

*2. The $\boldsymbol{\beta} = \boldsymbol{\gamma}$ case*

For $\beta = \gamma$, Eq. (91) yields

$$\Gamma(\infty) = \frac{m\omega_b^2}{2\pi \coth\left(\frac{\omega_b \hbar}{2k_B T}\right)} \frac{e^{-\frac{2V(x_b)}{\omega_b \hbar \coth\left(\frac{\omega_b \hbar}{2k_B T}\right)}}}{\sqrt{m^2\omega_b^2 - \gamma^2 \coth^2\left(\frac{\omega_b \hbar}{2k_B T}\right)}}. \tag{98}$$

and

$$\Gamma^{(T=0)}(\infty) = \frac{m\omega_b^2}{2\pi} \frac{e^{-\frac{2V(x_b)}{\omega_b \hbar}}}{\sqrt{m^2\omega_b^2 - \gamma^2}}. \tag{99}$$

A remarkable feature underlying our results is the absence of any damping constant $\gamma$ in the quantum Kramers rates (93), (95—97). This fact thereby suggests the existence of a quantum transport phenomenon coming about without dissipation (superfluidity) at thermal equilibrium *without violating* the quantum dissipation-fluctuation relation (54). This counter-intuitive quantum tunneling process in the strong friction regime shows up as a consequence of the arbitrary character of the parameter $\beta$ in the linear transformation (76) performed on our quantum Smoluchowski equation (74). That result is the same as the one obtained in Ref. [32] and has been already reported in studies about the influence of dissipation on the Landau—Zener transition at zero temperature [36].

It is worth pointing out that our non-dissipative quantum Kramers rate (93) can be defined at $T = 0$ in accordance with our previous work [32] and the Indian group's approach [19,20], but in contrast to Ankerhold's approach [13—16].

As far as the $\beta = \gamma$ case is concerned, our quantum tunneling escape rate (98) and (99) can depend upon the friction constant, thereby predicting no superfluidity phenomenon in the large friction realm. This feature seems to be in accordance with the works in Refs. [19,20, 37].

## C. The breakdown of the dissipation-fluctuation relation

Although our quantum Kramers escape rate (82) cannot be defined for $\mathcal{E}_\hbar(\infty) = 0$ without violating the dissipation-fluctuation relation, we may conjecture about the mathematical behavior of (82) as $\mathcal{E}_\hbar(\infty) \to 0$ corresponding to the extreme physical situation

$$\left|\frac{k\hbar^2}{8m\mathcal{E}_\hbar(\infty)}\right| \gg \left|\frac{\mathcal{E}_\hbar(\infty)}{2}\right|. \tag{100}$$

Under this condition the mean quantum-mechanical energy (52) approximately equals the mean kinetic energy, i.e.,

$$\langle \mathrm{E}(\infty)\rangle \cong \frac{k\hbar^2}{8m\mathcal{E}_\hbar(\infty)}. \tag{101}$$

From (82) in the limit $\mathcal{E}_\hbar(\infty) \to 0$, we obtain then the result

$$\Gamma(\infty) = c\delta(x_b). \tag{102}$$

The constant $c$ having dimensions such that $\Gamma(\infty)$ holds the dimensions of inverse of time. The non-exponential behavior of the escape rate (102) may suggest a breakdown of the dissipation-fluctuation relation provided that the condition (100) is valid, i.e., $\mathcal{E}_\hbar(\infty) \ll \omega_b\hbar/2$.

Experimental evidence for tunneling processes undergoing non-exponential decay have been reported in the literature in the context of isolated quantum systems [42]. Yet, our theoretical prediction (102) has to do with quantum open systems in the strong friction regime. So, we hope our outcome (102) could prompt experimental researches to investigate quantum tunneling processes in the presence of dissipative open systems in the large friction regime.

# V. SUMMARY AND OUTLOOK

In Sec. II we have derived the non-Markovian Smoluchowski equation (24) expressed in terms of the fluctuation-dissipation relation, given by the diffusion energy (18), and the averaging effects giving rise to the effective potential (25).

In Sec. III the dynamical quantization of our generalized Smoluchowski equation (27) for the case of a Brownian harmonic oscillator has been achieved. In so doing, our quantum phase-space Smoluchowski equation has been obtained [Eq.(40)] in terms of the quantum diffusion energy (35). In the long-time regime the quantum non-thermal stationary solution (49) has been derived.

As far as a quantum thermal bath is concerned, the quantum diffusion coefficient (55) associated with the quantum diffusion energy (54) has been found out. We have also obtained the steady thermal quantum probability distribution function (56). All our results are valid at zero temperature, $T = 0$. On the other hand, the Einstein's renowned diffusion constant (69), the energy equipartition (67), and the Maxwell—Boltzmann probability distribution function (70) have been rightly derived in the classical limit.

In Sec. IV the non-equilibrium quantum Kramers rate (82) for steady non-thermal open systems has been found on the basis of our quantum Smoluchowski equation (40). Our upshot (82) is expressed in terms of the dissipation-fluctuation relation, which is given by the steady quantum diffusion energy $\mathcal{E}_\hbar(\infty) = \gamma\mathcal{D}_\hbar(\infty)$, as well as in terms of the arbitrary parameter $\beta$ introduced by the linear transformation (76). We have also predicted a non-exponential behavior of our quantum escape rate (82) as $\mathcal{E}_{\hbar}(\infty) \to 0$, i.e., on the edge of the breakdown of the dissipation-fluctuation relation.

For thermal quantum systems our main finding is that the tunneling rate (88) may depend upon no friction constant [Eqs. (93,95—97)]. This fact hereby suggests the existence of a damping-independent tunneling process in the strong friction regime. Yet, supposing $\beta = \gamma$ in (91) leads to (98) and (99) in which the quantum tunneling phenomenon depends on the damping constant.

Lastly, it would be quite catchy to compare our quantum phase-space Smoluchowski equation (40) and its predictions with other accounts existing in the literature [13—25] and bring up new features underlying our non-Hamiltonian approach to open systems by applying it to the Brownian motion of particles dipped in fermionic and bosonic thermal reservoirs on the ground of our concept of quantum diffusion energy $\mathcal{E}_{\hbar}(\infty)$. This task will be achieved in a forthcoming paper [38].

## ACKNOWLEDGEMENTS

I wish to thank Professor Maria Carolina Nemes (Universidade Federal de Minas Gerais, Brazil) for the scientific support and Dr. Ping Ao (University of Washington) for bringing to my attention his papers present in Ref. [36]. I also acknowledge the Fundação de Amparo à Pesquisa do Estado de Minas Gerais (Fapemig) for the financial support.

## APPENDIX A: GENERALIZING THE SMOLUCHOWSKI EQUATION

The Langevin equation (13), given by

$$\frac{dX}{dt} = -\frac{1}{\gamma}\frac{dV}{dX} + b\Psi(t), \tag{A1}$$

gives rise to the Kolmogorov stochastic equation on configuration space [39—41]

$$\frac{\partial f(x,t)}{\partial t} = \mathbb{K} f(x,t), \tag{A2}$$

where the Kolmogorovian operator $\mathbb{K}$ acts upon the function $f(x,t)$ according to

$$\mathbb{K} f(x,t) = \sum_{k=1}^{\infty} \frac{(-1)^k}{k!}\frac{\partial^k}{\partial x^k}[A_k(x,t) f(x,t)], \tag{A3}$$

the coefficients $A_k(x,t)$ being given by

$$A_k(x,t) = \lim_{\varepsilon\to 0}\left[\frac{\langle(\Delta X)^k\rangle}{\varepsilon}\right], \tag{A4}$$

where the average values, $\langle(\Delta X)^k\rangle$, in the coefficients (A4) are to be calculated about the sharp values $q$, i .e.,

$$\mathcal{F}_{X\Psi}(x,\psi,t) = \delta(x-q)\mathcal{F}_{\Psi}(\psi,t). \tag{A5}$$

According to the Pawula theorem [40], if the coefficients $A_k(x,t)$ in (A4) are finite for every $k$ and if $A_k(x,t) = 0$ for some even $k$, then $A_k(x,t) = 0$ for all $k \geq 3$, thereby assuring the positivity of the probability density function $f(x,t)$. So, if $(x_2 - x_1)^3 \sim 0$ such that $(x_2 - x_1)^4 = 0$, it follows then that $A_k(x,t) = 0$, $k \geq 3$, where $x_2 = x(t+\varepsilon)$ and $x_1 = x(t)$. Accordingly, the Kolmogorov equation (A2) can be reduced to the generalized Smoluchowski equation (the so-called Fokker—Planck equation in configuration space)

$$\frac{\partial f(x,t)}{\partial t} = -\frac{\partial}{\partial x}[A_1(x,t)f(x,t)] + \frac{1}{2}\frac{\partial^2}{\partial x^2}[A_2(x,t)f(x,t)], \tag{A6}$$

where the drift coefficient is given by

$$A_1(x,t) = \lim_{\varepsilon\to 0}\left[\frac{\langle\Delta X\rangle}{\varepsilon}\right] = -\frac{1}{\gamma}\frac{dV}{dx} + b\langle\Psi(t)\rangle \tag{A7}$$

and the diffusion coefficient reads

$$A_2(x,t) = \lim_{\varepsilon\to 0}\left[\frac{\langle(\Delta X)^2\rangle}{\varepsilon}\right] = b^2 \lim_{\varepsilon\to 0}\frac{1}{\varepsilon}\iint_t^{t+\varepsilon} \langle\Psi(t')\Psi(t'')\rangle dt'dt'', \tag{A8}$$

with

$$\lim_{\varepsilon\to 0}\ \frac{1}{\varepsilon}\int_t^{t+\varepsilon} \langle\Psi(t')\rangle dt' = \langle\Psi(t)\rangle \tag{A9}$$

and

$$\lim_{\varepsilon\to 0}\int_t^{t+\varepsilon} \langle\Psi(t')\rangle dt' = 0. \tag{A10}$$

To evaluate the coefficients (A7) and (A8) we have used (A1) in the form

$$\Delta X \equiv X(t+\varepsilon) - X(t) = -\frac{\varepsilon}{\gamma}\frac{dV}{dX} + b\int_t^{t+\varepsilon} \Psi(t')dt'. \tag{A11}$$

## APPENDIX B: THE TIME-DEPENDENT DIFFUSION ENERGY

The time-dependent diffusion energy in (18)

$$\mathcal{E}(t) = \mathcal{E}(\infty) I(t) \tag{B1}$$

presents the correlational function

$$I(t) = \lim_{\varepsilon \to 0} \frac{1}{\varepsilon} \iint_{t}^{t+\varepsilon} \langle \Psi(t^{'}) \Psi(t^{''}) \rangle dt' dt'' \,. \tag{B2}$$

On the condition that the autocorrelation function, $\langle \Psi(t') \Psi(t'') \rangle$, can be given by

$$\langle \Psi(t') \Psi(t'') \rangle = \left( 1 - e^{-\frac{(t'+t'')}{2t_c}} \right) \delta(t' - t''), \tag{B3}$$

where $t_c$ is the correlation time of $\Psi(t)$ at times $t'$ and $t''$, it follows that (B2) becomes

$$I(t) = 1 - e^{-\frac{t}{t_c}}. \tag{B4}$$

Non-Markovian effects show up in the dimensionless function (B4) in view of the presence of the correlation time $t_c$, such that in the steady regime, $t \to \infty$, physically interpreted as $t \gg t_c$, it reduces to $I(\infty) = 1$, thus making the stochastic process Markovian.

# APPENDIX C: PROPERTIES OF THE QUANTUM FUNCTION (39)

The quantum function $W(x,p,t)$ defined in Eq. (39) does possess the following properties:

(i) just as $\rho(x,\eta,t)$ the function $W(x,p,t)$ is deemed to be real;

(ii) the classical limit provides

$$\lim_{\hbar\to 0} W(x,p,t) = f(x,t)\delta(p) = \mathcal{F}(x,p,t),$$

(iii) the normalization condition reads

$$\iint_{-\infty}^{\infty} W(x,p,t)dxdp = \iint_{-\infty}^{\infty} \rho(x,\eta,t)\delta(\eta)d\eta dx = \int_{-\infty}^{\infty} \rho(x,t)dx = 1;$$

(iv) in the classical limit the marginal distribution

$$\int_{-\infty}^{\infty} W(x,p,t)dp = \widetilde{W}(x,t)$$

leads to

$$\lim_{\hbar\to 0} \widetilde{W}(x,t) = f(x,t)\,;$$

(v) in quantum phase space the Fourier transformation (39) reinstates the stochastic character of the momentum $p$, i.e.,

$$\int_{-\infty}^{\infty} W(x,p,t)dx = W'(p,t).$$

(vi) In the classical limit, $\hbar \to 0$, we find the result

$$\lim_{\hbar\to 0} W'(p,t) = \lim_{\hbar\to 0} \int_{-\infty}^{\infty} W(x,p,t)dx = \delta(p),$$

meaning that in the classical domain the momentum turns out to be a deterministic variable in the large friction regime.

On the basis of the properties above, we could mathematically interpret the function $W(x, p, t)$ in (39) as a quasi-probability density, for it may assume negative values in some regions of quantum phase space $(x, p; \hbar)$. Even so, the average value of any quantum physical quantity $\mathcal{A}(X, P, t)$ can be deemed to be calculated as

$$\langle \mathcal{A}(X, P, t) \rangle = \iint_{-\infty}^{\infty} \mathcal{A}(x, p, t) W(x, p, t) dx dp.$$